# Mathematical simulation of package delivery optimization using a combination of carriers

Valentyn M. Yanchuk[1], Andrii G. Tkachuk[1], Dmitry S. Antoniuk[1], Tetiana A. Vakaliuk[1], and Anna A. Humeniuk[1]

1 Zhytomyr Polytechnic State University, Zhytomyr 10005, Ukraine

v.yanchuk@gmail.com, andru_tkachuk@ukr.net, dmitry_antonyuk@yahoo.com, tetianavakaliuk@gmail.com, gum_ann@ukr.net

## ABSTRACT

*A variety of goods and services in the contemporary world requires permanent improvement of services e-commerce platform performance. Modern society is so deeply integrated with mail deliveries, purchasing of goods and services online, that makes competition between service and good providers a key selection factor. As long as logistic, timely, and cost-effective delivery plays important part authors decided to analyze possible ways of improvements in the current field, especially for regions distantly located from popular distribution centers. Considering both: fast and lazy delivery the factor of costs is playing an important role for each end-user. Given work proposes a simulation that analyses the current cost of delivery for e-commerce orders in the context of delivery by the Supplier Fleet, World-Wide delivery service fleet, and possible vendor drop-ship and checks of the alternative ways can be used to minimize the costs. The main object of investigation is focused around mid and small businesses living far from big distribution centers (except edge cases like lighthouses, edge rocks with very limited accessibility) but actively using e-commerce solutions for daily activities fulfillment. Authors analyzed and proposed a solution for the problem of cost optimization for packages delivery for long-distance deliveries using a combination of paths delivered by supplier fleets, worldwide and local carriers. Data models and Add-ons of contemporary Enterprise Resource Planning systems were used, and additional development is proposed in the perspective of the flow selection change. The experiment is based on data sources of the United States companies using a wide range of carriers for delivery services and uses the data sources of the real companies; however, it applies repetitive simulations to analyze variances in obtained solutions.*

## KEYWORDS

*Simulation, Customer Behavior, Optimization, E-commerce.*

## 1. INTRODUCTION

### 1.1 Formulation of the problem.

It is very hard to imagine the contemporary world without planned deliveries, goods, and services ordered online or without scheduled charges for services, etc.

The delivery of e-commerce products has reached unexpected heights in the last few years. Most deliveries made with e-commerce consist of parcels, small packages, and food containers. Forrester Analytics builds the trends that the share of online retail will continue to grow steadily in the next years in the US [1]. Deliveries may have a variety of options like collection points, pickup locations, or direct delivery to the customer location.

From the perspective of distribution policy, there are two main types of deliveries from the perspective of the full path and last-mile logistics:

Fast delivery (within the same day or next day early morning), where the deadline for delivery passes very fast;

Lazy delivery (optional days for delivery) when the customer has to adapt to specific delivery days which is not necessary the next day or not even the nearest day or the week. Some Lazy deliveries can deliver on Wednesdays or Mondays and Fridays depending on the capacity of the delivery provider.

So, the workload of the online shop team is entirely packed. Every day the web-shop or any other online service operational employee collects items for an order then carefully packs them and sends via delivery service to the end-users. The great ease of this brings the online rate shopping tools that can calculate the costs precisely. With the variety of different vendor's carriers and modes of delivery, the end-user may choose between cost speed and flexibility.

Business to Business (B2B) E-commerce solutions have even more carriers and delivery modes due to the fact they are making delivery of smaller and bigger parcels sometimes even renting the place in a big car so-called Less than Truck Load (LTL) or Greater than Truck Load GTL services. Business to Consumer (B2C) segment has standardized.

At the market of the US, there are a lot of big players like FedEx, UPS, USPS, which makes the majority of deliveries for domestic, interstate, and international deliveries. All these carriers have web APIs that enable quick rate shopping for many e-commerce platforms [2]. At the same time, these big players are not picking bulky deliveries, which are greater than 65 kg, which can be an issue for car parts deliveries, etc. All these constraints are less impacting when you have specified add-ons that use online services helping rate-shop any basket or order and give almost an immediate result with the delivery rate per several shipping options. Nowadays this becomes rather standard to use online APIs that help quicker rate-shop the customers' basket and indicate the price. For better and precise calculation, the basket composition should also keep the delivery address, which should be validated by the delivery service to guarantee the parcel delivery. Dimensional packaging (width × height × depth) are optional but more important for bigger deliveries, which may be reviewed in other manuscripts, due to the different nature of the study.

Below is the simplified scheme of web-shop interaction with an online application integrated service, outlined by authors based on the online experience with the majority of online platforms.

(Step 1) the client forms basket at the online service.

(Step 2a) the basket is getting totals and tax calculated along with the total weight of the delivery (required).

(Step 2b) the dimensional information per product should be indicated per item for the possible use of packaging software calculating the costs for dimensional delivery (optional).

(Step 3) the Shipping origin (the depot or warehouse address) and Delivery address (in full, including Zip-code, city, state, country, Street, and Street 2 addresses should be provided (required).

(Step 4) Online API returns the calculated costs that can be added to the order total and prepared for payment with the full variety of methods and options for delivery.

(Step 5) Online ordering service forms the order for fulfilment and further this order is going for package. Some APIs provide reservation of the tracking number for the delivery, as long as this is finalized and will proceed to the carrier for delivery.

(Step 6) Delivery is scheduled and final delivery time is indicated to the client.

There is specifically highlighted the delivery address and shipping origin address, as they are the key factor for delivery costs calculation and dictate the distance or zone of delivery that plays an important role for the carriers.

Step 3 is an important part of the e-commerce flow, as this information should be dully validated for proper calculation. Besides, the Delivery Address should be recognized by the carrier to validate if there is a delivery to that address. In addition, here is another problem in the investigation: the rural addresses are hardly recognized by carriers, even if they are fully registered addresses with appropriate geolocation. Most carriers are covering specific network locations and points of delivery where the API can calculate the costs. Even the phone call to the carrier does not help much if the operator is seeing the same situation in his system.

There are always debates around the rural delivery areas [3], far distant location with limited delivery services [4] and the current research will not make the edge cases better, and however, the rural areas will certainly be considered. Unlike B2C, which is always dealing with different locations nowadays there are well developed rural zones and distant locations, where many farmers, smaller businesses concentrate their main locations, as these locations real estate and the land is cheaper. However, this does not resolve the problem of delivery, which becomes more and more problematic.

Several works already considered this problem [5] and many times the authors tried to streamline the delivery paths and wanted to go beyond their possibilities. For re-sellers it is important to save on costs as much as possible and keep the very good service. One of the costs, where the re-seller still loses is the transportation to the re-sellers' depots or packing points and then a calculation of the delivery should go further.

As known, the last-mile [6] delivery is currently regarded as one of the most expensive and least efficient portions of the entire supply chain.

This idea was observed in detail by Reyes and Taniguchi [7-8], based on the generalized vehicle routing problem (vendors' truck fleet in the given case) with time windows that have been explored. The study of Reyes, which also considered the earlier publications of urban and rural areas [9] proved that fleet of trucks type of delivery, could reduce total distances up to 40% for an application in the city of Atlanta. This research intends to build on the current state of the art by integrating the notion of travel time uncertainty.

As a very simple and direct solution the fleet of trucks for vendors cruising around the United States and delivery items, but the costs for truck maintenance and payments for the use of the service increases annually.

**1.2. Analysis of recent research and publications.**

The solution proposed is aimed mainly to optimize the e-commerce processes, help vendors and customers to get their orders in the best and cost-effective way. These intentions are highlighted in various publications directed to technical, economical sciences and a great deal of them still lies in the aspect of logistics optimization as well as forecasts of the upcoming infrastructural changes.

To cover this multidisciplinary approach let us disclose the existing relationships between e-commerce investigations made for shipments deliveries [3], including domains of domestic deliveries [4], rural areas deliveries [6-8], customer purchasing habits [11], simulation of deliveries in e-commerce systems.

The authors highlighted the approach of integration of delivery [2] were combined with the e-commerce solution with API services provided at the existing market. Overlapping of that work with publications of Routhier (2013) and Morganti and Dablanc (2014) uncovered the city and outside city delivery approach covering the transportation perspective and possible ways of further optimizations in that domain. Authors constantly suggested considering the direct and combined approaches of using the transportation system to optimize the time of delivery, however, the time is not always leading to a cost-effective solution.

Uwe Clausen, Christian Geiger (2016) in their hand-on testing of the last mile concept tried to cover the problems of building the optimal logistic way for larger vehicles using creating the Urban Consolidation Centers. However, this covers only the part of approaches that contemporary e-commerce systems have to consider, and smaller and medium enterprises may not arrange such centers on their own. Besides the approaches reviewed for Europe are not always easily applicable to the US with the higher distances and wider distribution of centers.

Reichheld and Schefter (2000), Abrham et al. (2015), Zelazny (2017) or Ehrenberger et al. (2015) observe that there is a significant relationship between long-term growth of companies' profitability and customer purchase intention [13] however that indicated a good insight that analyzing mid-size companies and trace the turnover and orders circulation it will be easier to identify the dependencies between the options people usually chooses and possible shipment options the current vendors can offer.

For testing data generators many researchers applied Monte Carlo methods, among them, are Sakas, Vlachos, (2014) to create simulations when the mean and variances are known – for our research we can use it to easier generate the experimental data for the current research. The authors decided to apply the Monte Carlo method based on recent researches described in this work as this suits best the nature of the research performed and reflects statistical inference regarding the original sampling domain. For the first step of the simulation, we have taken the original entries of orders accumulated in the test ERP and a very similar distribution will be generated with the Monte Carlo method. Comparison of Monte Carlo with the "what if" method gives better confidence intervals, that will give a better approximation.

## 2. THE METHOD

In a simplified view, the conceptual framework of the interaction of the re-sellers' eCommerce solution with vendors and delivery networks can be presented as in Figure 1.

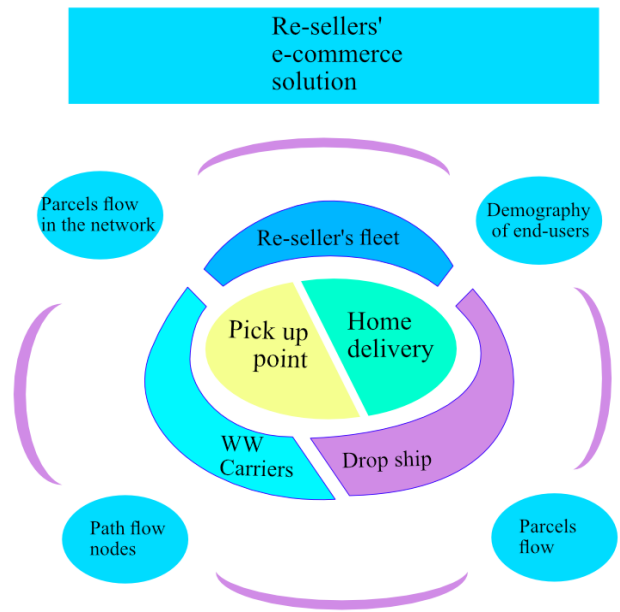

Figure. 1. The conceptual framework for a delivery network, from an operators' point of view. Adapted from [4]

So, in the current investigation, let us generate hundreds of orders based on real addresses of user form the Web-shop database and attempt to:

- Rate-shop those with World Wide Carriers (WW carriers) like UPS and FedEx with the basket compositions to export from sample shops.

- Select Pick-Up delivery from the nearest location (according to [10-11] it still gives a lot of benefits from an economic point of view).

- Use the truck fleet of the vendor for distant deliveries with the truck cost taken per ride.

Allocate an option to deliver via partner network, which will be the fleet of the supplier, which anyways deliver the parcels to the online sellers. Drop-ship service is also having costs, but these costs are covered with the Dealer Discount rate, which can be still considered as a benefit. For the given task and simplicity let us set up the discount on delivery costs to 5%.

For a given investigation of the products selected is not that important, as at Step 2a important is the weight of the basket assembled. To mitigate the risk that clients will not be satisfied with generated basket fetching their order – authors filtered out orders that are having too high prices for delivery since that is obvious the customers may give up delivering them.

Dimensional values for products are not important, the focus is at the selection of the delivery per weights and the range of weight will be between 60, which still falls into International deliveries and already interested in LTL service providers and are both good for Own and Supplier's truck fleets.

For the experiment taken 4 US companies where the vendor provides the goods for re-seller, who sell the products to the customer. Having statistical information on these companies the authors can compare the results of simulations with orders that were found and compared in the system via scanning the database and comparing the costs incurred to date. The companies are selected in pares to investigate 2 fleets of re-sellers versus the fleet of vendors for delivery of drop-ship orders.

The analysis of the results will be presented in the context of how much costs the customer saves with the time window not more than 1 day longer than the vendor's truck fleet, which is a guaranteed way of delivery.

## 3. RESULTS

Using the Monte Carlo method there were generated n orders in 4 networks located in California and distributed across 4 networks S1, R2, R3, and S4 that have the following distribution of clients per with urban and rural zones and percentage of network with the distant rural zone (see the Table 1).

Table 1. Distribution of the network with rural and urban zones

| Netw | Classification | Urban zone clients | Rural zone clients | Orders generated for a week | % of Rural distant zone orders |
|---|---|---|---|---|---|
| S1 | Supplier of R2 | 210 | 380 | 1200 | 30 |
| R2 | Re-seller of S1 and S4 | 250 | 320 | 1800 | 25 |
| R3 | Re-seller of S4 | 200 | 180 | 1300 | 45 |
| S4 | Supplier of R1 and R2 | 320 | 260 | 1500 | 26 |

There is no overlapping of clients between networks; however, dependencies of Supplier and Re-seller are given in the column Network Dependencies and orders distributions.

It has been performed 8 simulations (by 2 runs for each type of simulation to calculate the deviation) using Test Dynamics NAV (Enterprise Resource Planning system) custom developed add-on feeding the system with Delivery addresses of a given distribution of Rural Distant zone orders and called the Online and network services for the price calculation. A custom add-on is developed based on Retail Add-on that can trace the system of discounts and particularly evaluate the behavior of the customer. The ease of use of this Add-on got rid of behavioral model validation, as Retail Add-on collects the customer orders in the context of basket composition and the selected shipping method.

Method Monte-Carlo is used to check if the number of the rural distant zone is more than 10% deviation from setting up original parameters as recommended in [11].

When the Monte-Carlo method was applied, the following distribution of the carriers was made across networks (see Table2).

Comparison of orders generated by the Monte-Carlo method shown in the last column of the table.

Table 2. Distribution of the selection for Far rural zone

| Carriers | Average price per order with the same weight | S1 | R2 | R3 | S4 | % of deviation |
|---|---|---|---|---|---|---|
| FedEx | 120 USD | 219 | 73 | 41 | - | 12 |
| Vendor Truck | 80 USD | 301 | 523 | 162 | 400 | 18 |
| Supplier Truck | 75 USD | 490 | 18 | 212 | 5 | 20 |
| PickUP location | 20 USD* | 130 | 186 | 395 | 265 | 38 |
| Total | - | 1140 | 800 | 810 | 670 | - |

*The price per pick-up appeared due to the fact of overdue for taking orders from location center

Including into simulation the factor of cancellation of pick-up locations (distributors may potentially cancel delivery to a specific location) and see how the average price impacts the customer if they keep selecting the rest of vendors. The first run shows that customers start selecting either Supplier truck or FedEx, especially if the weight combination comes closer to the FedEx edge weight of 60 kg (See Table 3).

Table 3. Second simulation results overview

| Carriers | Average price per order with the same weight | S1 | R2 | R3 | S4 | % of deviation |
|---|---|---|---|---|---|---|
| FedEx | 100 USD | 393 | 57 | 432 | 260 | 22 |
| Vendor Truck | 80 USD | 17 | 48 | 65 | 24 | 18 |
| Supplier Truck | 75 USD | 730 | 695 | 497 | 386 | 15 |
| Total | - | 1140 | 800 | 810 | 670 | |

Results were reviewed with the descent analysis of the simulation results and, due to the fact the ERP generation add-on is used, the dependency of the Vendor Truck, which delivers to the pick-up location of Supplier organization, that chain was filtered.

To avoid gaps in the behavioral model it was decided to export data from Google Analytics Extended (supplied by Google Tag Manager) choice preferences versus a list of available vendors and repeat the simulation. It was found that the generation add-on is trying to mimic the web-shop user's behavior in case of absence of Pickup location following the tendency of the user's choice.

As an additional limitation, it was decided to exclude FedEx. At this moment of simulation, it became easier to trace that customers' orders became in the status of Dropped, as some users indeed were leaving web-shops not finding the guaranteed delivery provider. According to [12] the variety of delivery methods for the B2C segment should have a wider range of prices for delivery, then the choice of the customer will be either in favor or cheapest or most reliable carrier (personal user preference). However, B2B users in this simulation behaved exactly like B2C customers, preferring to abandon the basket or order and leave, rather than select the cheapest option. However, another tendency was noticed, that abandoned baskets were noticed from the users, who reached to the limit of the price they are ready to pay; thus, the factor of the economy played an important role. The last attempt of simulation taken with the assumption that the customer may use the Vendor's truck for the same price as the supplier, but the additional day may be added to the total route time (see Table IV).

Table 4. Forth simulation results overview

| Carriers | Average price per order with the same weight | S1 | R2 | R3 | S4 | % of deviation |
|---|---|---|---|---|---|---|
| Vendor Truck | 75 USD | 621 | 729 | 549 | 493 | 20 |
| Supplier Truck | 75 USD | 519 | 71 | 261 | 177 | 22 |
| Total | - | 1140 | 800 | 810 | 670 | |

As it is seen from the results the clients behaved more reluctant to the delay in 1 day, however, they selected the home delivery via the Vendor's truck, rather than the supplier. I assume this happened due to the price change and most probably, no competence with the supplier appeared a key factor for Vendors' truck network. This also proves the statements described in [12] that users ordering bigger deliveries are usually more reluctant to have guaranteed delivery, rather than have it fast, but the tendency is equivalent for both: B2B and B2C segments as it is also indicated in [13].

The current investigation is complete, however during the investigation, it was identified, that simulation of orders creation with the only ERP data without analytical data will only indicate the quantitative part, without the qualitative part that can be filled with additional statistical data per customer on personal preferences of choices in different cases of orders created in the live systems and presence or absence of the carrier in certain circumstances. Thus, adding Google Analytics Extended data on user choice allowed us to complete the investigation.

Google Analytics covered the validation of Users Acceptance Testing (UAT) and reflection of their behavior for generated orders. Additionally, the use of Google Analytics given an insight into the comparison of generated orders and directed to UAT for different groups of users during the application testing and run-down tests performed in all 4 companies.

Table 5, indicating the coverage of orders by comparison of generated orders and directed to UAT with orders recently submitted in the system (completed orders) will serve as the validation baseline for the approach suggested.

Table 5. Coverage of the real orders for Far rural zones* that correspond to generated orders per simulation

| Carriers | Average price per order with the same weight | S1, % | R2, % | R3, % | S4, % |
|---|---|---|---|---|---|
| FedEx | 120 USD | 92 | 92 | 86 | 75 |
| Vendor Truck | 80 USD | 90 | 87 | 95 | 65 |
| Supplier Truck | 75 USD | 89 | 68 | 73 | 85 |
| PickUP location | 20 USD* | 76 | 64 | 86 | 92 |
| Average | - | 87.65 | 77.75 | 85 | 79.25 |

* Far Rural zones are distantly located areas having lower delivery capacities, mostly implemented by private delivery networks and lower flows of international carriers.

As we see, generated orders have rather high coverage from the data perspective and lie in the frame of deviations we calculated recently (Table 6.).

Table 6. Coverage of the real orders for Domestic zones that correspond to generated orders per simulation

| Carriers | Average price per order with the same weight | S1, % | R2, % | R3, % | S4, % |
|---|---|---|---|---|---|
| FedEx | 120 USD | 97 | 92 | 95 | 97 |
| Vendor Truck | 80 USD | 90 | 92 | 90 | 96 |
| Supplier Truck | 75 USD | 95 | 88 | 89 | 87 |
| PickUP location | 20 USD* | 90 | 89 | 86 | 92 |
| Average | - | 93 | 90.25 | 90 | 93 |

A higher level of coverage for orders generated for domestic zones is relatively easy to explain from the perspective of a higher number of orders generated for domestic zones.

**4. Conclusions**

The series of simulations performed given a possibility to evaluate the behavior of users in case of absence of usual carriers and taking the decision of cheaper solution It also demonstrates options the web-shop owner may offer to the mid and small-size businesses alternative delivery services, where the drop-ship of vendor delivers to the final destination or drop-point with minimal costs if the additional discount is given to keep the order fulfillment. As a continuation of this work can be developed and on-line web-service to support delivery network visualizing deliveries of vendors and re-sellers, where all packages can be loaded into the truck, correct the route, and deliver goods and services with Vendor's fleet, as long as the route and discount permits.

To apply the current solution to the real industrial case will involve the expansion of the Vendor and Re-seller networks sharing the same data model for orders and street validation. In the case of contemporary ERP systems use that should not be a very difficult problem, however, it may involve additional 3-rd party services.

The paper highlighted the combination of approaches of drop-ship deliveries and selection of different routes for total costs optimization and indicated that behavior of the client is not always driven by economic circumstances, but also by habits and tendencies of customers, so well described in [7].

For future investigations, authors will involve Google Tag Manager data, collected for customers, as the import of options for checkout selection helped discover how customers may potentially abandon basket if they do not see the option of the habitual carrier or the price for the order versus delivery cost exceeds a specific limit.

**Authors**

**Valentyn Yanchuk**, Associate Professor at the Department of Automation and Computer-Integrated Technologies named after Prof. B.B. Samotkin, Zhytomyr Polytechnic State University, Zhytomyr, Ukraine

Valentyn Yanchuk, born in 1975, received a Candidate of Technical Sciences degree (Ph.D.) from the Pukhov Institute for Modelling in Energy Engineering, National Academy of Sciences of Ukraine (IMPE) in 2002. Since 1997, he has been working in the field of Software Engineering and Information Technologies at the Zhytomyr Polytechnic State University, where he is currently working at the position of Associate Professor and in the practical field of IT business. His research interests include software engineering, business analysis, computer-based modeling, E-Commerce. He has published several papers and proceedings in the journals and material of conferences.

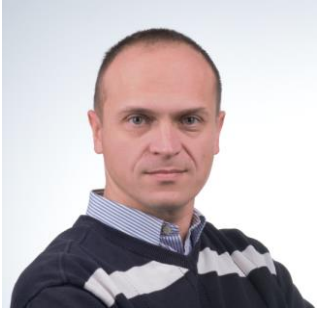

WWW: https://ztu.edu.ua
E-mail: v.yanchuk@gmail.com

Assoc. Prof. **Andrii Tkachuk**, Head of the Department of Automation and Computer-Integrated Technologies named after prof. B.B. Samotokin, Zhytomyr Polytechnic State University, Zhytomyr, Ukraine.

Andrii Tkachuk, born in 1989, received a Candidate of Technical Sciences degree from the National Technical University of Ukraine "Kyiv Polytechnic Institute", Ukraine, in 2014. Since 2012, he has been working in the field of information technologies, automation, and robotics at the Zhytomyr Polytechnic State University. Andrii Tkachuk is an Expert of the Scientific Council of the Ministry of Education and Science of Ukraine, a Board member of NGO "Youth integration center". His research interests include information technologies, automated aviation gravimetric systems, mobile robotics, armament stabilization systems. He has published several papers in international journals, is a reviewer of The scientific journal Aviation which is included in the Scopus database.

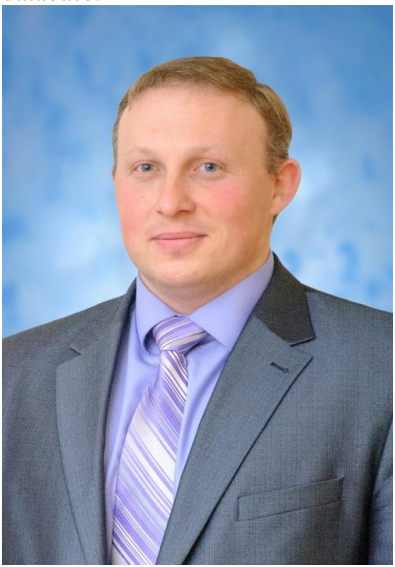

Www:
https://ztu.edu.ua/ua/structure/faculties/fikt/kakt.php
e-mail: andru_tkachuk@ukr.net

**Dmytro Antoniuk**, Assistant Professor of the Department of Software Engineering, Zhytomyr Polytechnic State University, Zhytomyr, Ukraine

Dmytro Antoniuk, born in 1981, received a Candidate of Pedagogical Sciences degree (Ph.D.) from the Institute of Information Technologies and Learning Tools, Ukraine, in 2018. Since 2003, he has been working in the field of Software Engineering and Information Technologies at the Zhytomyr Polytechnic State University, where he is currently an Assistant Professor of the Department of Software Engineering and in the practical field of IT business. His research interests include software engineering, business in IT, computer-based business-simulation, economic and financial literacy of technical professionals. He has published several papers and proceedings in the journals and material of conferences.

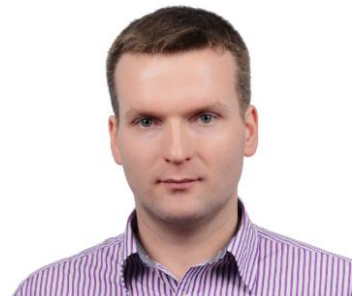

WWW: https://ztu.edu.ua
E-mail: Dmitry_antonyuk@yahoo.com

Dr. **Tetiana Vakaliuk**, professor of the Department of Software Engineering, Zhytomyr Polytechnic State University, Zhytomyr, Ukraine.

Tetiana Vakaliuk, born in 1983, received a Candidate of Pedagogical Sciences degree from the National Pedagogical Dragomanov University, Ukraine, in 2013, and a Doctor of Pedagogical Sciences degree from the Institute of Information Technologies and Learning Tools of the National Academy of Sciences of Ukraine, in 2019. Since 2019, she has been working in the field of information technologies at the Zhytomyr Polytechnic State University. Her research interests include information technologies, ICT in Education, Cloud technologies. She has published several papers in international journals, is a member of editorial boards of Information Technologies and Learning Tools, Zhytomyr Ivan Franko State University Journal: Pedagogical Sciences, Collection of Scientific Papers of Uman State Pedagogical University.

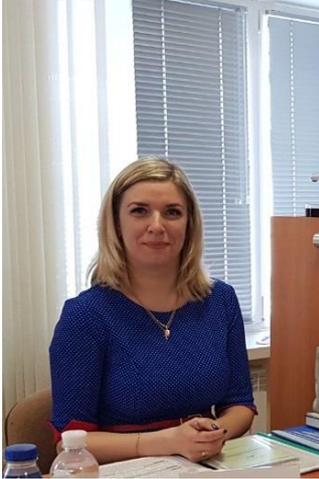

WWW: https://sites.google.com/view/neota
e-mail: tetianavakaliuk@gmail.com

**Anna Humeniuk** Associate Professor at the Department of Automation and Computer-Integrated Technologies named after Prof. B.B. Samotokin, Zhytomyr Polytechnic State University, Zhytomyr, Ukraine

Anna Humeniuk, born in 1986, received a Candidate of Technical Sciences degree (Ph.D.) from the National Technical University of Ukraine "Igor Sikorsky Kyiv Polytechnic Institute" in 2011.

Her scientific interests include automated measuring systems, gravimetry, flexible manufacturing system design automation.

She has published several papers and proceedings in the journals and material of conferences.

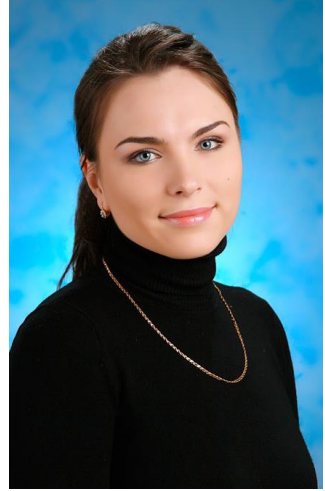

WWW: https://ztu.edu.ua
E-mail: gum_ann@ukr.net